\documentclass{ws-procs975x65}

\begin{document}

\title{COLLAPSE AND EVAPORATION OF A CANONICAL\\ SELF-GRAVITATING GAS}

\author{Cl\'ement SIRE and Pierre-Henri CHAVANIS}

\address{Laboratoire de Physique Th\'eorique -- IRSAMC,
Universit\'e de Toulouse III \& CNRS,\\
Toulouse, FR-31062, France\\
E-mail: clement.sire@irsamc.ups-tlse.fr and chavanis@irsamc.ups-tlse.fr
\\Web: www.lpt.ups-tlse.fr}

\begin{abstract}
We review the out-of-equilibrium properties of a self-gravitating gas
of particles in the presence of a strong friction and a random force (canonical
gas). We assume a bare diffusion coefficient of the form $D(\rho)=T\rho^{1/n}$,
where $\rho$ is the local particle density, so that the equation of state is
$P(\rho)=D(\rho)\rho$. Depending on the spatial dimension $d$, the index $n$,
the temperature $T$, and whether the system is confined to a finite box or not,
the system can reach an equilibrium state, collapse or evaporate.
This article focuses on the latter cases, presenting a complete dynamical phase
diagram of the system.
\end{abstract}


\bodymatter

\section{Introduction and background}

There is currently a renewed interest in the study of long-range interacting
systems, particularly outside the realm of astrophysics. These systems may exhibit
inequivalence of statistical ensembles (\textit{e.g.}
canonical \textit{vs} microcanonical) which affects their equilibrium properties
(possibility of negative specific heat in the microcanonical ensemble) and even
more dramatically, their dynamical properties. In the context of self-gravitating
systems, the Newtonian dynamics is too complicated to permit an exhaustive
analytical treatment. Hence, in the past few years, we have developed a model
of self-gravitating particles in the presence of a strong friction and
a random force \cite{col1}, for which inertial effects are negligible. The
main interest of this model is to be analytically tractable in many situations,
while presenting dynamical phases reminiscent of their Newtonian
counterparts. In addition, this model is intimately related to the Keller-Segel
model of bacterial chemotaxis \cite{col2}.

We thus consider particles obeying the equations of motion
$\frac{d{\bf x}_i}{dt}=-\nabla \Phi+\sqrt{2D}{\bf \eta}_i,$
where ${\bf \eta}_i$ is a delta correlated random Gaussian force, and
$\Phi$ is the  gravitational potential. In a proper mean-field limit, which
becomes exact for an infinite number of particles, the density obeys
a Fokker-Planck (or Smoluchowski) equation coupled to the Poisson equation
(the gravitational constant is set equal to $G=1$):
\begin{equation}
{\frac{\partial\rho}{\partial t}}=\nabla \cdot (\nabla P+\rho\nabla\Phi),\quad
\Delta\Phi=\rho.
\label{dim1}
\end{equation}
The pressure is related to the diffusion coefficient by $P(\rho)=D(\rho)\rho$, and
the  isothermal case corresponds to $D=T$, while we will consider here the more
general polytropic case  $D=T\rho^{1/n}$ \cite{col3}. In addition, the system of
total mass $M$ can be placed in a spherical bounded domain of radius $R$ or
in an unbounded space.

In the context of chemotaxis \cite{col2}, $\rho$ is the density of bacteria,
and $c=-\Phi$ is the  density of a chemical that they secrete. The bacteria are
attracted by the regions of high density of this chemical which generates an
effective long-range interaction between them. This interaction exactly takes the form of
gravity when neglecting the diffusion and the degradation of the chemical.

Although we shall see that the actual phase diagram depends crucially on the
value of the index  $n$, the general physics is controlled by the parameter $T$
which we assimilate to the temperature. When $T$ is small, the kinetic
pressure is not strong enough to compensate gravity and the system may collapse.
For large $T$, the system is at equilibrium in a bounded domain or evaporates in
an unbounded domain. In the next section, we present the main results that we
have obtained in \cite{col1,col2,col3,col4,col5,col6,virial,col7} concerning these
different dynamical phases.

\section{Collapse and post-collapse dynamics and the critical index $n_*$}

Defining the critical index $n_*=\frac{d}{d-2}$, the system can be shown to
always have a  polytropic equilibrium state for $0<n<n_*$, whether the system is
confined or not. Hence we shall concentrate on dynamical properties of the case
$n>n_*$ and $n<0$ (including the isothermal case $n=\infty$ ($d>2$)), and on the
interesting particular case $n=n_*$ (including the isothermal case $n=\infty$
($d=2$)).

\subsection{$n>n_*$ and the isothermal case $n=\infty$ ($d>2$)}

We have shown \cite{col1,col3,col4,col5} that in a bounded domain, and for
$T<T_c$, the system displays a finite time singularity at which the central
density diverges as $\rho_0(t)\sim (t_{\rm coll}-t)^{-1}$. The density profile
takes a scale-invariant form $\rho(r,t)=\rho_0(t)f[r/r_0(t)]$, where the core
radius $r_0(t)$ is related to the central density by $r_0^2(t)={T/\rho_0^{1-1/n}(t)}$.
The density scaling function decays as $f(x)\sim x^{-\alpha}$, with
$\alpha=\frac{2n}{n-1}$, and can be analytically obtained in the case $n=\infty$
($d>2$) \cite{col1}. In this important case, $t_{\rm coll}$  was also computed
exactly close to $T_c$ \cite{col5}.

For $t>t_{\rm coll}$, during what we dubbed the post-collapse regime \cite{col4},
a delta peak condensate grows at the center as $M_0\sim(t-t_{\rm
coll})^{d(n-1)/(2n)-1}$, and ultimately saturates exponentially to the total
mass $M$, with a rate which was computed analytically for low $T$, by using
semi-classical quantum methods \cite{col4}. Meanwhile, the residual density
obeys a reverse scaling of the form $\rho_{\rm
res.}(r,t)=\rho_r(t)f_r[r/r_r(t)]$,
where $\rho_r(t)\sim (t-t_{\rm coll})^{-1}$, and $f_r$ decays with the same
exponent $\alpha=\frac{2n}{n-1}$ as $f$.

In an unbounded domain\cite{col6}, the gas collapses below a \textit{non
universal} critical temperature $T_c$ depending on the initial state, or
evaporates for $T>T_c$. In this evaporation regime, gravity becomes
asymptotically negligible and free diffusion is observed, with sub-corrections
due to gravity which have been obtained analytically \cite{virial,col6}.

\subsection{$n<0$}
The case of negative $n$ \cite{col7}, including the logotropic case $n=-1$, is
similar to the case treated above for $n<-d/2$, with a collapse scaling
exponent $\alpha=\frac{2n}{n-1}$. However, for $-d/2<n<0$, the collapse is
controlled by the $T=0$ fixed point \cite{col1,col3,col7} with
$\alpha=\frac{2d}{d+2}$.

\subsection{$n=n_*$ and the isothermal case $n=\infty$ ($d=2$)}

In this critical case \cite{col1,virial,col6}, $T_c$ is independent of the radius $R$
of the confining box ($T_c=M/4$, in $d=2$). For $T<T_c$, there is still a finite
time singularity but a delta peak condensate of mass
$M_c=(T/T_c)^{d/2}M$ forms at the center, at $t=t_{\rm coll}$. Moreover, the
residual density takes a pseudo scale-invariant form with an effective scaling
exponent $\alpha(t)$ very slowly saturating to $\alpha=d$, as $t$ goes to $t_{\rm
coll}$. In an unbounded domain, the gas collapses below the \textit{universal} critical
temperature $T_c$, or evaporates for $T>T_c$.
In this evaporation regime, the scaling density profile is strongly affected by
gravity and the gas expansion is such that $\langle r^d(t)\rangle \sim (T-T_c)t$
\cite{virial,col6}.

\section{Summary}

In Table~\ref{Table1}, we summarize the static and dynamic phase diagram of a
self-gravitating gas of particles with a bare diffusion coefficient
$D=T\rho^{1/n}$, where $T$ is the temperature and $d$ the spatial dimension.
This table illustrates the crucial role played by the critical index
$n_*=\frac{d}{d-2}$.
\begin{table}
\centering
\tbl{Static and dynamic phase diagram of a self-gravitating gas.\hfill{ }}
{
\tiny
\begin{tabular}{@{}|c|c|c|c|@{}}
\Hline
& & & \\
   \textbf{{Index $n$}} & \textbf{{Temperature}}  & \textbf{{Bounded domain}}  &  \textbf{{Unbounded domain}}   \\
& & & \\
 \Hline
   &     & Metastable equilibrium state & $\bullet$ Evaporation : \\
     &  $T>T_c$    &  (local minimum of free energy): & asymptotically free normal   \\
$n=\infty$ ($d>2$)    &   & box-confined isothermal sphere   & diffusion (gravity negligible)\\
  \cline{2-3}
     &     & Self-similar collapse with $\alpha=2$  & $\bullet$ Collapse:   \\
   &   $T<T_c$   & and self-similar post-collapse leading & pre-collapse  and post-collapse \\
  &  &  to a Dirac peak of mass $M$  &  as in a bounded domain  \\
\hline
    &  $T>T_c$     & Equilibrium state:  & Equilibrium state:  \\
 $0<n<n_*$     &     & box-confined (incomplete) polytrope  & complete polytrope  \\
\cline{2-3}
      & $T<T_c$     & Equilibrium state:  & (compact support)  \\
   &     & complete polytrope (compact support) &    \\
\hline
    &      & Metastable equilibrium state & $\bullet$ Evaporation:  \\
     &   $T>T_c$  &  (local minimum of free energy): & asymptotically free anomalous   \\
 $n_*<n<\infty$     &     &  box-confined polytropic sphere &  diffusion (gravity negligible)    \\
\cline{2-3}
 $n<0$     &    & Self-similar collapse with $\alpha=2n/(n-1)$ & $\bullet$ Collapse:   \\
   &   $T<T_c$    &  or $\alpha=2d/(d+2)$ ($-d/2<n<0$) and   & pre-collapse and  post-collapse \\
   &     &    post-collapse leading to a Dirac peak & as in a bounded domain  \\
\hline
 & $T>T_c$     & Equilibrium state: & Self-similar evaporation \\
     &    &  box-confined (incomplete) polytrope  & modified by self-gravity  \\
\cline{2-4}
     &      & Pseudo self-similar collapse & \\
 $n=n_*$   &     & leading to a Dirac peak of & Collapse \\
   &   $T<T_c$  & mass $(T/T_c)^{d/2}M$ $+$ halo. & as in a bounded domain \\
   &     & This is followed by a post-collapse &  \\
   &     & leading to a Dirac peak of mass $M$ &  \\
\cline{2-4}
     & $T=T_c$ & Infinite family of steady states & As in a bounded domain \\
\Hline
\end{tabular}}
\label{Table1}
\end{table}
\bibliographystyle{ws-procs975x65}

\end{document}